\documentstyle[11pt,amssymb,times]{article}
\newcommand{\be}{\begin{equation}}
\newcommand{\ee}{\end{equation}}
\newcommand{\ba}{\begin{array}}
\newcommand{\ea}{\end{array}}
\newcommand{\bea}{\begin{eqnarray}}
\newcommand{\eea}{\end{eqnarray}}

\newcommand{\p}{\partial}
\newcommand{\ol}{\overline}
\newcommand{\ti}{\tilde}
\newcommand{\la}{\langle}
\newcommand{\ra}{\rangle}
\renewcommand{\l}{\newline\null}

\def\hbar{h\!\!\!/}
\begin{document}
\begin{titlepage}
May 1997\hfill PAR-LPTHE 97/18
\vskip 5cm
\begin{center}
{\bf 
TWO RESULTS CONCERNING \boldmath{$CP$} VIOLATION FOR \boldmath{$J=0$} MESONS.}
\end{center}
\vskip .5cm
\centerline{B. Machet
     \footnote[1]{Member of `Centre National de la Recherche Scientifique'.}
     \footnote[2]{E-mail: machet@lpthe.jussieu.fr.}
           }
\vskip 5mm
\centerline{{\em Laboratoire de Physique Th\'eorique et Hautes Energies,}
     \footnote[3]{LPTHE tour 16\,/\,1$^{er}\!$ \'etage,
          Universit\'e P. et M. Curie, BP 126, 4 place Jussieu,
          F 75252 PARIS CEDEX 05 (France).}
}
\centerline{\em Universit\'es Pierre et Marie Curie (Paris 6) et Denis
Diderot (Paris 7);} \centerline{\em Unit\'e associ\'ee au CNRS URA 280.}
\vskip 1.5cm
{\bf Abstract:} I show, in the framework of a $SU(2)_L \times U(1)$
gauge theory  for $J=0$ mesons expressed as scalar or pseudoscalar
di-quark fields, that:\l
-\ the existence, at the fermionic level, of a complex mixing matrix of the 
Kobayashi-Maskawa type is not a sufficient condition for 
electroweak mass eigenstates to be different from $CP$ eigenstates;\l
-\ unitarity constrains  this phenomenon to arise from an admixture of states 
with different parities, and present experiments probing ``indirect'' 
$CP$ violation are likely to be interpreted with $P$ violation only.
\smallskip

\vfill
\end{titlepage}
\section{Introduction.}

The only yet observed phenomenon of $CP$ violation
\cite{ChristensonCroninFitchTurlay} \cite{Argus}, called ``indirect'' 
$CP$ violation \cite{Nir}, is that some electroweak mass eigenstates 
are not $CP$ eigenstates.
The unavoidable presence, for a number of generations $N/2 \geq 3$,
of a complex number among the entries of the quark mixing matrix 
\cite{KobayashiMaskawa} is the preferred mechanism to  trigger it
in the framework of a $SU(2)_L \times U(1)$ electroweak gauge theory for
quarks \cite{GlashowSalamWeinberg}. Other possibilities need enlarging 
the scalar sector of the model \cite{LeeWeinberg}.

We shall deal here with mesons only, and stick to their interpretation as
composite di-quark fields, first proposed by Gell-Mann \cite{GellMann} for
the case of three flavours; the success of the $SU(3)$ classification 
of the corresponding eigenstates was next extended to $N$ flavours,
with a similar role played by the diagonal subgroup of the chiral 
$U(N)_L \times U(N)_R$ group.
The importance of the latter and of its breaking, specially as far as strong 
interactions are concerned, was put forward long ago \cite{CurrentAlgebra};
we shall refer to the corresponding eigenstates as
the ``flavour'' or ``strong'' eigenstates (strong interactions are
considered to be flavour independent).
 
The quarks being in the fundamental representation of $U(N)$, one is naturally 
led to consider mesons as $N \times N$ matrices \cite{Machet1}; 
each of them is given in addition a quantum number
$(+1)$ or $(-1)$ when acted upon by the parity-changing operator $\cal P$,
such that their total number $(2N^2)$ of degrees of freedom matches the one
of scalar and pseudoscalar $J=0$ mesons.
The action on them of the $U(N) \times U(N)$ generators,
which are $N \times N$ matrices, too, is defined inside the associative
algebra that they form. Fermions can then be forgotten, though the group action 
as defined in \cite{Machet1}
can be easily recovered by acting with the chiral group on both fermionic
components of the mesonic wave function and introducing the appropriate 
``left'' and ``right'' projectors, with a $\gamma_5$ matrix, 
respectively for the generators of $U(N)_L$ and $U(N)_R$.

The extension of the Glashow-Salam-Weinberg model \cite{GlashowSalamWeinberg} 
to $J=0$ mesons that I proposed in \cite{Machet1} is 
thus a $SU(2)_L \times U(1)$ gauge theory of matrices. As the action of
the gauge group can only be defined if its generators are also $N \times N$
matrices, it is considered as a subgroup of the chiral group. Its
orientation within the latter has to be compatible with the customary action
of the electroweak group on fermions, and is determined by a unitary $N/2
\times N/2$ matrix which is nothing else than the Cabibbo-Kobayashi-Maskawa 
mixing matrix $\Bbb K$ \cite{Cabibbo} \cite{KobayashiMaskawa}.

The $SU(2)_L$ generators are
\footnote{This construction of course requires an even number $N$ of flavours.}
\be
{\Bbb T}^3_L = {1\over 2}\left(\begin{array}{rrr}
                        {\Bbb I} & \vline & 0\\
                        \hline
                        0 & \vline & -{\Bbb I}
\end{array}\right),\
{\Bbb T}^+_L =           \left(\begin{array}{ccc}
                        0 & \vline & {\Bbb K}\\
                        \hline
                        0 & \vline & 0           \end{array}\right),\
{\Bbb T}^-_L =           \left(\begin{array}{ccc}
                        0 & \vline & 0\\
                        \hline
                        {\Bbb K}^\dagger & \vline & 0
\end{array}\right),
\label{eq:SU2L}
\ee
and act trivially on the $N$-vector of quarks
\be 
\Psi =
\left( 
\ba{c}  u\\ c\\ \vdots \\d\\ s\\ \vdots \ea 
\right).
\label{eq:psi}\ee
$\Bbb I$ is the $N/2 \times N/2$ identity matrix.

The $U(1)$ generator satisfies the Gell-Mann-Nishijima relation (written in
its ``chiral'' form)
\be
({\Bbb Y}_L,{\Bbb Y}_R) = ({\Bbb Q}_L,{\Bbb Q}_R) - ({\Bbb T}^3_L,0),
\label{eq:GMN}\ee
and the customary electric charge operator
\be
{\Bbb Q} = \left(\begin{array}{ccc}
                        2/3 & \vline & 0\cr
                        \hline
                        0 & \vline & -1/3
           \end{array}\right),
\ee
yields back the usual expressions for the ``left'' and ``right'' hypercharges
\be
{\Bbb Y}_L = {1\over 6}{\Bbb I}, \quad {\Bbb Y}_R = {\Bbb Q}_R.
\ee
$\Bbb Q$ turns out to be the ``third'' generator of the custodial $SU(2)_V$
symmetry uncovered in \cite{Machet1}.

The electroweak eigenstates can be classified into two types of 
quadruplets, respectively 
``even'' and ``odd'' by the parity changing operator $\cal P$. Both write

\vbox{
\bea
& &\Phi(\Bbb D)=
({\Bbb M}\,^0, {\Bbb M}^3, {\Bbb M}^+, {\Bbb M}^-)(\Bbb D)\cr =
& & \left[
 {1\over \sqrt{2}}\left(\begin{array}{ccc}
                        {\Bbb D} & \vline & 0\\
                        \hline
                        0 & \vline & {\Bbb K}^\dagger\,{\Bbb D}\,{\Bbb K}
                   \end{array}\right),
{i\over \sqrt{2}} \left(\begin{array}{ccc}
                        {\Bbb D} & \vline & 0\\
                        \hline
                        0 & \vline & -{\Bbb K}^\dagger\,{\Bbb D}\,{\Bbb K}
                   \end{array}\right),
i\left(\begin{array}{ccc}
                        0 & \vline & {\Bbb D}\,{\Bbb K}\\
                        \hline
                        0 & \vline & 0           \end{array}\right),
i\left(\begin{array}{ccc}
                        0 & \vline & 0\\
                        \hline
                        {\Bbb K}^\dagger\,{\Bbb D} & \vline & 0
                    \end{array}\right)
             \right],\cr
& &
\label{eq:reps}
\eea
}
where $\Bbb D$ is a real $N/2 \times N/2$ matrix.
That the entries ${\Bbb M}^+$ and ${\Bbb M}^-$ are, up to a sign,
hermitian conjugate ({\em i.e.} charge conjugate) requires that the $\Bbb
D$'s are restricted to symmetric or antisymmetric matrices.
Because of the presence of an ``$i$'' for the for ${\Bbb M}^{3,\pm}$ and not
for ${\Bbb M}^0$, the quadruplets always mix entries of different behaviour
by hermitian (charge) conjugation, and are consequently not hermitian
representations.

Each of them is  the sum of two
doublets of $SU(2)_L$, and also the sum of one singlet plus one triplet of
the custodial diagonal $SU(2)_V$. The $\cal P$-even and $\cal P$-odd
quadruplets do not transform in the same way
by $SU(2)_L$ (the Latin indices $i,j,k$ run from $1$ to $3$); 
for ${\cal P}$-even quadruplets, one has
\bea
{\Bbb T}^i_L\,.\,{\Bbb M}^j_{{\cal P}even} &=& -{i\over 2}\left(
              \epsilon_{ijk} {\Bbb M}^k_{{\cal P}even} +
                           \delta_{ij} {\Bbb M}_{{\cal P}even}^0
                              \right),\cr
{\Bbb T}^i_L\,.\,{\Bbb M}_{{\cal P}even}^0 &=&
                                {i\over 2}\; {\Bbb M}_{{\cal P}even}^i;
\label{eq:actioneven}
\eea
while ${\cal P}$-odd quadruplets transform according to
\bea
{\Bbb T}^i_L\,.\,{\Bbb M}_{{\cal P}odd}^j &=& -{i\over 2}\left(
                   \epsilon_{ijk} {\Bbb M}_{{\cal P}odd}^k -
                           \delta_{ij} {\Bbb M}_{{\cal P}odd}^0
                              \right),\cr
{\Bbb T}^i_L\,.\,{\Bbb M}_{{\cal P}odd}^0 &=&
                        \hskip 5mm  -{i\over 2}\; {\Bbb M}_{{\cal P}odd}^i,
\label{eq:actionodd}
\eea
and only representations transforming alike, $\cal P$-even or $\cal P$-odd, 
can be linearly mixed.
The (diagonal) charge operator acts indifferently on both types of
representations by:
\bea
{\Bbb Q}\,.\,{\Bbb M}^i &=& -i\,\epsilon_{ij3} {\Bbb M}^j,\cr
{\Bbb Q}\,.\,{\Bbb M}^0 &=& 0.
\label{eq:chargeaction}
\eea
The misalignment of ``strong'' and electroweak eigenstates, resulting from
the one of the electroweak group with respect to the chiral group, is
conspicuous from the presence of the mixing matrix in the definition
(\ref{eq:reps}).

By adding or subtracting eqs.~(\ref{eq:actioneven}) and (\ref{eq:actionodd}), 
and defining scalar ($\Bbb S$) and pseudoscalar ($\Bbb P$) fields by
\be
({\Bbb M}_{{\cal P}even} + {\Bbb M}_{{\cal P}odd}) = {\Bbb S},
\label{eq:scalar}
\ee
and
\be
({\Bbb M}_{{\cal P}even} - {\Bbb M}_{{\cal P}odd}) = {\Bbb P},
\label{eq:pseudo}
\ee
one finds two new types of stable quadruplets which include objects of different
parities, but which now correspond to a given $CP$ quantum number, depending
in particular whether $\Bbb D$ is a symmetric or skew-symmetric matrix
\cite{Machet1}
\be
({\Bbb M}\,^0, \vec {\Bbb M}) = ({\Bbb S}^0, \vec {\Bbb P}),
\label{eq:SP}
\ee
and
\be
({\Bbb M}\,^0, \vec {\Bbb M}) = ({\Bbb P}\,^0, \vec {\Bbb S});
\label{eq:PS}
\ee
they transform in the same way by the gauge group, according to 
eq.~(\ref{eq:actioneven}), and thus can be linearly mixed.  As they span the 
whole space of $J=0$ mesons too, this last property makes them
specially convenient to build an electroweak gauge theory.

Taking the hermitian conjugate of any representation $\Phi$ swaps the
relative sign between ${\Bbb M}^0$ and $\vec{\Bbb M}$; as a consequence,
$\Phi^\dagger_{{\cal P}even}$ transforms by $SU(2)_L$
as would formally do a ${\cal P}$-odd representation, and vice-versa;
on the other hand, the quadruplets (\ref{eq:reps}) are also representations
of of $SU(2)_R$, the action of which is obtained by swapping 
eqs.~(\ref{eq:actioneven}) and (\ref{eq:actionodd}) \cite{Machet1};
so, the hermitian conjugate of a given representation of $SU(2)_L$ is a 
representation of $SU(2)_R$ with the same law of transformation, and 
vice-versa. The same result holds for any (complex) linear
representation $U$ of quadruplets transforming alike by the gauge group.

The link with usually defined $J=0$ ``strong'' mesonic eigenstates proceeds
as follows: consider for example the case $N=4$, for which $\Bbb K$ shrinks
back to the Cabibbo mixing matrix; the pseudoscalar $\pi^+$ meson is 
represented in our notation, up to a scaling factor (see below), by the matrix
\be
\Pi^+ = \left( \ba{rrcrr}   &  &\vline & 1 &  0 \\
                                 &  &\vline & 0 &  0 \\
                            \hline
                                 &  &\vline &   &     \\
                                 &  &\vline &   &  \ea \right),
\ee
since, sandwiched between two 4-vectors $\Psi$ of quarks (\ref{eq:psi}),
it gives
\be
\ol\Psi\ \Pi^+\ \Psi  = \bar u d,
\ee
which indeed corresponds, according to the classification by flavour $SU(4)$,
to the $(+1)$ charged pion. One identifies similarly the other strong
pseudoscalar mesons, for example  $K^+ = \bar u s$, $D^+ = \bar c d$, $D_s^+ =
\bar c s$.  So, for example, with the scaling that has to be introduced 
(see \cite{Machet2} \cite{Machet3} \cite{Machet1}, where I
show that it leads in particular to the correct leptonic decays), 
the pseudoscalar entry ${\Bbb P}^+$ with charge $(+1)$ 
\be
{\Bbb P}^+ =
i \left(\ba{rrcrr}   &  &\vline & c_\theta &  s_\theta \\
                             &  &\vline &-s_\theta &  c_\theta \\
                            \hline
                             &  &\vline &   &     \\
                             &  &\vline &   &  \ea \right),
\ee
corresponding to the matrix
\be
{\Bbb D}_1 = \left( \ba{cc} 1 & 0 \\
                            0 & 1 \ea \right),
\ee
represents the following linear combination of pseudoscalar mesons
\be
{\Bbb P}^+({\Bbb D}_1) = i{f\over \la H\ra}\left(c_\theta (\pi^+ + D_s^+) +
                                         s_\theta (K^+ -D^+)\right),
\ee
where $f$ is the leptonic decay constant of the mesons, that we consider to
be the same for all of them,  and $H$ is the Higgs boson 
(see the remark at the end of Appendix A).

\section{Quadratic invariants.}

To every representation is associated a quadratic expression invariant
by the electroweak gauge group $SU(2)_L \times U(1)$
\be
{\cal I} = ({\Bbb M}^0, \vec {\Bbb M})\otimes ({\Bbb M}^0, \vec {\Bbb M})=
 {\Bbb {\Bbb M}}\,^0 \otimes {\Bbb {\Bbb M}}\,^0 +
                 \vec {\Bbb M} \otimes \vec {\Bbb M};
\label{eq:invar}
\ee
the ``$\otimes$'' product is a tensor product, not
the usual multiplication of matrices and means the product
of fields as functions of space-time; $\vec {\Bbb M} \otimes \vec {\Bbb M}$
stands for $\sum_{i=1,2,3} {\Bbb M}\,^i \otimes  {\Bbb M}\,^i$.

For the relevant cases $N=2,4,6$, there exists a set of $\Bbb D$ matrices 
(see appendix A) such that the algebraic sum (specified below) 
of invariants extended over all  representations defined by 
(\ref{eq:SP},\ref{eq:PS},\ref{eq:reps})
\bea
&&
{1\over 2} 
\left((\sum_{symmetric\ {\Bbb D}} - \sum_{skew-symmetric\ {\Bbb D}})
\left( ({\Bbb S}^0, \vec {\Bbb P})({\Bbb D})
                     \otimes  ({\Bbb S}^0, \vec {\Bbb P})({\Bbb D})
- ({\Bbb P}^0, \vec {\Bbb S})({\Bbb D})
                     \otimes  ({\Bbb P}^0, \vec {\Bbb S})({\Bbb D})
\right)\right)\cr
&=&
{1\over 4}
\left((\sum_{symmetric\ {\Bbb D}} - \sum_{skew-symmetric\ {\Bbb D}})
\left(\Phi_{{\cal P}even}({\Bbb D})\otimes\Phi^\dagger_{{\cal P}odd}({\Bbb D})
  +\Phi_{{\cal P}odd}({\Bbb D})\otimes \Phi^\dagger_{{\cal P}even}({\Bbb D})
\right) \right)\cr
&&
\label{eq:Idiag}\eea
is diagonal both in the electroweak basis and in the basis of 
strong eigenstates:
in the latter basis, all terms are normalized alike to $(+1)$
(including the sign).
Note that two ``$-$'' signs  occur in eq.~(\ref{eq:Idiag})
\footnote{Eq.~(\ref{eq:Idiag} specifies eq.~(25) of
\cite{Machet1}, in which the ``$-$'' signs were not explicitly written.}
:\l
- the first between the $({\Bbb P}^0, \vec{\Bbb S})$ and 
$({\Bbb S}^0, \vec{\Bbb P})$ quadruplets, because, as seen on
eq.~(\ref{eq:reps}), the ${\Bbb P}^0$ entry of the former has no ``$i$''
factor, while the $\vec{\Bbb P}$'s of the latter do have one; as we define 
all pseudoscalars without an ``$i$'' 
(like $\pi^+ = \bar u d$), a $(\pm i)$ relative factor has to be introduced
between the two types of representations, yielding a ``$-$'' sign in
eq.~(\ref{eq:Idiag});\l
- the second for  the representations corresponding to skew-symmetric 
$\Bbb D$ matrices, which have an opposite
behaviour by charge conjugation ({\em i.e.} hermitian conjugation)
as compared to the ones with symmetric ${\Bbb D}$'s.

The kinetic part of the $SU(2)_L \times U(1)$ Lagrangian for $J=0$ mesons
is built from the same combination (\ref{eq:Idiag})
 of invariants, now used for the covariant 
derivatives of the fields with respect to the gauge group;
it is thus diagonal in both the strong and electroweak basis, too.

Other invariants can be built like tensor products of two representations 
transforming alike by the gauge group: two $\cal P$-odd or two $\cal P$-even, 
two $({\Bbb S}^0,\vec {\Bbb P})$, two $({\Bbb P}^0,\vec {\Bbb S})$, or one
$({\Bbb S}^0,\vec {\Bbb P})$ and one $({\Bbb P}^0,\vec {\Bbb S})$; for example
such is
\be
{\cal I}_{1\ti 2} = ({\Bbb S}^0,\vec {\Bbb P})({\Bbb D}_1) \otimes 
                ({\Bbb P}^0,\vec {\Bbb S})({\Bbb D}_2)
 ={\Bbb S}^0({\Bbb D}_1) \otimes {\Bbb P}^0({\Bbb D}_2) +
                  \vec {\Bbb P}({\Bbb D}_1) \otimes \vec {\Bbb S}({\Bbb D}_2).
\label{eq:I12}\ee
According to the remark made in the previous section, all the above expressions 
are also invariant by the action of $SU(2)_R$.

They naturally enter the mass terms in the Lagrangian, and there are
{\em a priori} as many $(N^2/2)$ independent mass scales as there are 
independent representations. Introduced in a gauge invariant way, they share 
with the leptonic case the same arbitrariness; the ratios of mesonic masses 
have here the same status as the one between the muon and the electron.  
Note that we have given a purely electroweak origin to
the mass splittings, since, from the diagonalization property of
eq.~(\ref{eq:Idiag}), equal electroweak mass terms also
correspond to equal mass terms for strong eigenstates.

\subsection{The basic property of the quadratic invariants.}

The quadratic $SU(2)_L$ invariants are not {\em a priori} self conjugate 
expressions
\footnote{The hermitian combination (\ref{eq:Idiag}), used to build the 
kinetic terms, is special in this respect too.}
and have consequently  no definite property by hermitian conjugation;
in particular, the one associated with a given representation $U$ is 
$U \otimes U$ and {\em not} $U \otimes U^\dagger$ (we have seen in the
previous section that $U$ and $U^\dagger$ do not transform alike
by the gauge group).

As far as one only deals with representations of the type of
eqs.~(\ref{eq:reps},\ref{eq:SP},\ref{eq:PS}), it has no consequence since 
each of their entries has a well defined behaviour by hermitian conjugation:
the associated quadratic invariants are then always hermitian. 
But electroweak mass eigenstates are in general  (complex) linear
combinations of them with, consequently, no definite behaviour by hermitian 
(charge) conjugation.
%

\section{Two results concerning \boldmath{$CP$} violation.}

Let us use the invariants associated to the $N^2/4$ quadruplets (\ref{eq:SP}) 
and $N^2/4$ quadruplets (\ref{eq:PS}), which all transform by 
(\ref{eq:actioneven}), to construct a $SU(2)_L \times U(1)$ gauge Lagrangian 
for the $2N^2$ scalar and pseudoscalar $J=0$ mesons.

\subsection{A first result.}

Unitarity compels this Lagrangian to be hermitian, in particular its quadratic 
part.

Suppose that it has been diagonalized and let us restrict for the sake of 
simplicity to a subsystem of two non-degenerate electroweak mass eigenstates
$U$ and $V$; they are in general complex linear combinations of quadruplets
(\ref{eq:SP}) and (\ref{eq:PS}), and transform by $SU(2)_L$ 
according to (\ref{eq:actioneven}). $\cal L$ writes, for example
\be
{\cal L} = {1\over 2}(\p_\mu U\otimes \p^\mu U -\p_\mu V\otimes \p^\mu V
 - m_U^2 U\otimes U + m_V^2 V\otimes V +\cdots).
\ee
with $m_U^2 \not = m_V^2$.

Hermiticity  yields the two following  equations, 
coming respectively from the kinetic and mass terms
\be\left\{\ba{l}
(U\otimes U - V\otimes V)^\dagger = U\otimes U - V\otimes V,\cr
(m_U^2 U\otimes U - m_V^2 V\otimes V)^\dagger =m_U^2 U\otimes U - m_V^2
V\otimes V,
\ea\right. \ee
which, if we reject complex values of the (mass)$^2$, entail
\be U = \pm U^\dagger,\quad V=\pm V^\dagger;
\ee
unitarity thus requires  that the electroweak mass eigenstates be also
$C$ eigenstates.

Consequence: {\em if electroweak mass eigenstates are observed not to be $CP$
eigenstates, they can only be  mixtures of states with different parities.}

\subsection{A second result.}

Suppose that we have a complex mixing matrix $\Bbb K$; the following
Lagrangian for $J=0$ mesons, where the sum is extended to all
representations defined by eqs.~(\ref{eq:SP},\ref{eq:PS},\ref{eq:reps}),
is nevertheless hermitian,
($D_\mu$ is the covariant derivative with respect to $SU(2)_L \times U(1)$)

\vbox{
\bea
{\cal L}= &&{1\over 2}\sum_{symmetric\ {\Bbb D}}
           \left(D_\mu ({\Bbb S}^0, \vec {\Bbb P})(\Bbb D)
                             \otimes D^\mu ({\Bbb S}^0, \vec {\Bbb P})(\Bbb D) 
          - m_D^2 ({\Bbb S}^0, \vec {\Bbb P})(\Bbb D)
                             \otimes ({\Bbb S}^0, \vec {\Bbb P})(\Bbb D)
        \right.\cr
&&\hphantom{{1\over 2}\sum_{symmetric\ {\Bbb D}}}
\left. -\left(D_\mu ({\Bbb P}^0, \vec {\Bbb S})(\Bbb D)
                    \otimes D^\mu ({\Bbb P}^0, \vec {\Bbb S})(\Bbb D) 
          - \ti m_D^2 ({\Bbb P}^0, \vec {\Bbb S})(\Bbb D)
                    \otimes ({\Bbb P}^0, \vec {\Bbb S})(\Bbb D)
\right) \right)\cr
- &&{1\over 2}\sum_{skew-symmetric\ {\Bbb D}}
\left(D_\mu ({\Bbb S}^0, \vec {\Bbb P})(\Bbb D)
                     \otimes D^\mu ({\Bbb S}^0, \vec {\Bbb P})(\Bbb D) 
          - m_D^2 ({\Bbb S}^0, \vec {\Bbb P})(\Bbb D)
                     \otimes ({\Bbb S}^0, \vec {\Bbb P})(\Bbb D) \right.\cr
&&\hphantom{{1\over 2}\sum_{symmetric\ {\Bbb D}}}
        \left.-\left(D_\mu ({\Bbb P}^0, \vec {\Bbb S})(\Bbb D)
                     \otimes D^\mu ({\Bbb P}^0, \vec {\Bbb S})(\Bbb D) 
          - \ti m_D^2 ({\Bbb P}^0, \vec {\Bbb S})(\Bbb D)
                \otimes ({\Bbb P}^0, \vec {\Bbb S})(\Bbb D)\right)\right),\cr
& &
\eea
}
and its mass eigenstates, being the $({\Bbb S}^0, \vec {\Bbb P})$ and
$({\Bbb P}^0, \vec {\Bbb S})$ representations given by 
(\ref{eq:SP},\ref{eq:PS}) are $CP$ eigenstates \cite{Machet1}.
It is of course straightforward to also build hermitian $SU(2)_L \times
U(1)$ invariant quartic terms.

Consequence: {\em The existence of a complex phase in the mixing matrix for
quarks is not a sufficient condition for the existence of electroweak mass
eigenstates for $J=0$ mesons different from $CP$ eigenstates}.

\section{Conclusion.}

Until we observe direct $CP$ violation \cite{Nir}, and if we stick to a 
$SU(2)_L \times
U(1)$ gauge theory of particles, the origin of observed features of $CP$
violation for $J=0$ mesons transforming like composite di-quark fields 
by the chiral group $U(N)_L \times U(N)_R$ should be looked for into a 
mixture of scalar and pseudoscalar states, and be interpreted as a simple 
effect of parity violation at the mesonic level.
\bigskip
\begin{em}
\end{em}
\newpage\null
{\Large\bf Appendix}

\appendix

\section{Diagonalizing eq.~(\protect\ref{eq:Idiag}) in the basis of strong
eigenstates: a choice of \boldmath{$\Bbb D$} matrices.}

The property is most simply verified for the ``non-rotated'' 
$SU(2)_L \times U(1)$ group and representations \cite{Machet1}.

\subsection{$N=2.$}

Trivial case: $\Bbb D$ is a number.

\subsection{$N=4.$}

The four $2\times 2$ $\Bbb D$ matrices ($3$ symmetric and $1$
skew-symmetric) can be taken as
\be
{\Bbb D}_1 = \left( \ba{cc} 1 & 0 \cr
                            0 & 1     \ea \right),\ 
{\Bbb D}_2 = \left( \ba{rr} 1 & 0 \cr
                            0 & -1    \ea \right),\ 
{\Bbb D}_3 = \left( \ba{cc} 0 & 1 \cr
                            1 & 0     \ea \right),\ 
{\Bbb D}_4 = \left( \ba{rr} 0 & 1 \cr
                           -1 & 0     \ea \right).
\ee

\subsection{$N=6.$}

The nine $3 \times 3$ $\Bbb D$ matrices ($6$ symmetric and $3$ skew-symmetric),
can be taken as
\bea
& &{\Bbb D}_1 = \sqrt{{2\over 3}}\left( \ba{ccc}
                                1  &  0  &  0  \cr
                                0  &  1  &  0  \cr
                                0  &  0  &  1 \ea \right), \cr
& &{\Bbb D}_2 ={2\over\sqrt{3}} \left( \ba{ccc}
                \sin\alpha  &     0    &    0    \cr
       0     & \sin(\alpha\pm{2\pi\over 3})&   0   \cr
       0     &                    0        & \sin(\alpha\mp{2\pi\over 3})
       \ea\right),\ 
{\Bbb D}_3 ={2\over\sqrt{3}} \left( \ba{ccc}
                \cos\alpha  &     0    &    0    \cr
       0     & \cos(\alpha\pm{2\pi\over 3})&   0   \cr
       0     &                    0        & \cos(\alpha\mp{2\pi\over 3})
       \ea\right), \cr
& & {\Bbb D}_4 =\left( \ba{ccc}
                                0  &  0  &  1 \cr
                                0  &  0  &  0 \cr
                                1  &  0  &  0   \ea \right),\ 
     {\Bbb D}_5 =\left( \ba{rrr}
                                0  &  0  &  1 \cr
                                0  &  0  &  0 \cr
                               -1  &  0  &  0   \ea \right),\cr
& & {\Bbb D}_6 = \left( \ba{ccc}
                                0  &  1  &  0  \cr
                                1  &  0  &  0  \cr
                                0  &  0  &  0   \ea \right), \ 
 {\Bbb D}_7 = \left( \ba{rrr}
                                0  &  1  &  0  \cr
                               -1  &  0  &  0  \cr
                                0  &  0  &  0   \ea \right), \ 
 {\Bbb D}_8 = \left( \ba{ccc}
                                0  &  0  &  0  \cr
                                0  &  0  &  1  \cr
                                0  &  1  &  0   \ea \right), \ 
 {\Bbb D}_9 = \left( \ba{rrr}
                                0  &  0  &  0  \cr
                                0  &  0  &  1  \cr
                                0  & -1  &  0   \ea \right),\cr
& &
\eea
where $\alpha$ is an arbitrary  phase.
 
{\em Remark}: as ${\Bbb D}_1$ is the only matrix with a non vanishing trace,
${\Bbb S}^0({\Bbb D}_1)$ is the only neutral scalar matrix with the
same property;  we take it as the Higgs boson.  

Considering that it is the only scalar with a non-vanishing vacuum 
expectation value prevents the occurrence of a hierarchy problem 
\cite{GildenerWeinberg}.

This last property is tantamount, in the ``quark language'', to taking
the same value for all condensates $\la\bar q_i q_i\ra, i=1 \cdots N$, 
in agreement with the flavour independence of ``strong interactions''
between fermions, supposedly at the origin of this phenomenon in the 
traditional framework.

As the spectrum of mesons is, in the present model, disconnected from a
hierarchy between quark condensates (see section 2),
it is not affected by our choice of a single Higgs boson.

\newpage\null
\begin{em}

\end{em}

\end{document}